\documentclass[prd,aps,showpacs]{revtex4}

\usepackage{amsmath}
\usepackage{amssymb}
\usepackage{latexsym}
\usepackage{amsfonts}
\usepackage{epsfig}
\usepackage{psfrag}
\usepackage{graphicx}

\newcommand{\bea}{\begin{eqnarray}}
\newcommand{\ea}{\end{eqnarray}}
\newcommand{\eea}{\end{eqnarray}}

\begin{document}

\title{A  Gauge-Gravity Relation in the One-loop Effective Action} 

\author{G\"ok\c ce Ba\c sar and  Gerald V. Dunne}
\affiliation{ Department of Physics, University of Connecticut, 
Storrs CT 06269, USA}

\begin{abstract}
We identify an unusual new gauge-gravity relation: the one-loop effective action for a massive spinor in 2n dimensional AdS space is expressed in terms of  precisely the same function [a certain multiple gamma function] as the one-loop effective action for a massive charged scalar in 4n dimensions in a maximally symmetric background electromagnetic field [one for which the eigenvalues of $F_{\mu\nu}$ are maximally degenerate, corresponding in 4 dimensions to a self-dual field, equivalently to a field of definite helicity], subject to the identification $F^2\leftrightarrow \Lambda$, where $\Lambda$ is the gravitational curvature. Since these effective actions generate the low energy limit of all one-loop multi-leg graviton or gauge amplitudes, this implies a nontrivial gauge-gravity relation at the non-perturbative level and at the amplitude level.
\end{abstract}



\maketitle

Remarkable progress has been made recently in relating perturbative graviton scattering amplitudes with gauge scattering amplitudes \cite{Bern:2002kj}. The basic correspondence can be expressed figuratively as
\begin{eqnarray}
\left({\rm gravity\,\, amplitude}\right) \sim \left({\rm gauge\,\, amplitude}\right)^2
\label{gg}
\end{eqnarray}
where $\sim$ means that the correspondence relates terms in the $\epsilon$ expansion with dimensional regularization in $d=(4-2\epsilon)$ dimensions, and the various external indices must be treated appropriately \cite{Dixon:1996wi}. This surprising correspondence has a natural origin in string theory, based on the  Kawai-Lewellen-Tye (KLT) relations \cite{Kawai:1985xq,Bern:1999ji,BjerrumBohr:2003vy} between amplitudes for open and closed string theories. These relations can be projected to the field theory limit, resulting in new quantum field theoretic identifications \cite{Bern:2002kj}. The factorization property can also be seen directly in basic quantum field theory language by studying the factorization properties of tree level graviton scattering amplitudes into photon scattering amplitudes \cite{Berends:1988zp,Choi:1994ax,Holstein:2006ry}. In this note we point out another possible perspective on these correspondences at the one-loop level, based on the effective action, which is the generating function for {\it all} one-loop amplitudes, with any number of external photon or graviton lines. As such, our observation is in the spirit of early analyses from the string theory perspective \cite{Metsaev:1986yb,Tseytlin:1986ti}, but it is based on simple field theory results that are very well-known.

The most compact and  explicit expressions for effective actions arise for external background fields having constant gauge or gravitational curvature \cite{gvd,Candelas:1975du,Dowker:1975xj,Allen:1983dg,Candelas:1983ae,Burges:1985qq,Avramidi:2000bm}, which therefore generate low-energy limits of amplitudes \cite{Itzykson:1980rh,Martin:2003gb}. Further simplifications arise when the constant curvature has maximal symmetry, so we consider this case in order to illustrate our point. Thus, on the gauge theory side  we are led to consider the quantization of massive fields in a gauge background of constant field strength and maximal symmetry. In $d=4$ this corresponds to a constant self-dual field strength \cite{duff-isham,Dubovikov:1981bf,BialynickiBirula:1981ym,Dunne:2002qf}, which means external photon lines of definite helicity. (For {\it massless} loop particles, such amplitudes are well known to have dramatically simple forms, best understood in the spinor helicity and twistor formalisms \cite{Bern:2002kj,Dixon:1996wi}; but remarkable simplifications occur also for massive particles when the background field has maximal symmetry \cite{Dunne:2002qf}). On the gravity side, it is natural to consider the quantization of massive particles in a constant curvature anti-de Sitter space \cite{Avis:1977yn,Allen:1985wd,Burgess:1984ti,Camporesi:1992tm}. The relevant effective actions are well known, on both the gauge and gravity side, having been computed many times, using a variety of different approaches. Here we  point out an unusual relation between these two types of effective action, and interpret this as a manifestation of the gauge-gravity relation (\ref{gg}) at a non-perturbative level in addition to the perturbative scattering amplitude level.

On the gauge theory side, consider a charged scalar field $\phi$ in $d$-dimensional Euclidean  space, with $d$ even [eventually we will actually take $d=4n$], coupled to a background abelian gauge field with maximally symmetric constant field strength. Thus, the $d\times d$  antisymmetric matrix $F_{\mu\nu}$ can be written in block diagonal form
\begin{eqnarray}
F_{\mu\nu}=f\, {\rm diag}_{d/2}\left(
\begin{pmatrix}
0&1\cr
-1&0
\end{pmatrix}
\right)
\label{f}
\end{eqnarray}
with the single parameter $f$ characterizing the strength of the background field.
The spectral problem factorizes into $d/2$ copies of the two dimensional Landau level problem, so  
the position space propagator is simply expressed in proper-time form in terms of the heat kernel \cite{gvd}:
\begin{eqnarray}
G(x, x^\prime )&=&\int_0^\infty \frac{ds}{(4\pi s)^{d/2}}\, e^{-m^2 s}\,\left(\frac{f\, s}{\sinh(f\, s)}\right)^{d/2} \,  e^{-\frac{f |x-x^\prime|^2}{4}\coth(fs)}\nonumber\\
&=&\frac{1}{2f}\left(\frac{f}{2\pi}\right)^{d/2}\Gamma \left(\frac{d}{4}+\frac{m^2}{2f}\right)U\left(\frac{d}{4}+\frac{m^2}{2f}, \, \frac{d}{2};\, \frac{f |x-x^\prime|^2}{2}\right) e^{-\frac{f |x-x^\prime|^2}{4}}
\label{g1}
\end{eqnarray}
where $U(a,b;\,z)$ is the confluent hypergeometric function. For the effective action, we need the coincident-point propagator
\begin{eqnarray}
G(x, x)&=&\left(\frac{f}{4\pi}\right)^{d/2}\int_0^\infty ds\, \frac{e^{-m^2 s}}{\sinh^{d/2}(f\, s)}\nonumber\\
&=& \frac{1}{4\pi}\left(\frac{f}{4\pi}\right)^{d/2-1}\int_0^1 du\, u^{-d/2}\left(1-u\right)^{\frac{d}{4}+\frac{m^2}{2f}-1} 
\, \left(1+u\right)^{\frac{d}{4}-\frac{m^2}{2f}-1} \nonumber\\
&=& \frac{2^{d/2-1}}{f}\left(\frac{f}{4\pi}\right)^{d/2} \frac{\Gamma \left(1-\frac{d}{2}\right)\Gamma \left(\frac{d}{4}+\frac{m^2}{2f}\right)}{\Gamma\left(1-\frac{d}{4}+\frac{m^2}{2f}\right)}
\label{g}
\end{eqnarray}
where in the second line we have substituted $u=\tanh(f s)$, and these expressions are to be understood in the sense of dimensional regularization, as explained below.

On the gravity side,  consider a spinor field of mass $m$ in $d$ dimensional anti-de Sitter  (AdS) space, with curvature $\Lambda$, represented by its Euclidean section  $H^d$, with $d$ even. Again, the effective action is well known, and can be simply derived from the position space propagator \cite{Allen:1985wd,Burgess:1984ti,Camporesi:1992tm,odintsov}. 
For the effective action, we need the coincident-point propagator \cite{Allen:1985wd,Burgess:1984ti,Camporesi:1992tm,odintsov}: 
\begin{eqnarray}
{\mathcal D}(x, x)&=&-\frac{m}{\Lambda} (2\Lambda)^{d/2} \int_0^\infty \frac{ds}{(4\pi s)^{d/2}} U\left(1+\frac{m}{\sqrt{\Lambda}}, 1+\frac{d}{2}; \frac{1}{s}\right) s^{-1-m/\sqrt{\Lambda}}\nonumber\\
&=&-\frac{1}{\sqrt{\Lambda}}\left(\frac{\Lambda}{2\pi}\right)^{d/2}\frac{\Gamma \left(1+\frac{m}{\sqrt{\Lambda}}\right)\Gamma \left(\frac{d}{2}+\frac{m}{\sqrt{\Lambda}}\right)}{\Gamma \left(1+\frac{2m}{\sqrt{\Lambda}}\right)} \,_2F_1\left(\frac{d}{2}+\frac{m}{\sqrt{\Lambda}} ,\,\frac{m}{\sqrt{\Lambda}} ;\,1+\frac{2m}{\sqrt{\Lambda}} ;\,1 \right)\nonumber\\
&=&-\frac{1}{\sqrt{\Lambda}}\left(\frac{\Lambda}{2\pi}\right)^{d/2}\frac{\Gamma\left(1-\frac{d}{2}\right)\Gamma\left(\frac{d}{2}+\frac{m}{\sqrt{\Lambda}}\right)}{\Gamma\left(1-\frac{d}{2}+\frac{m}{\sqrt{\Lambda}}\right)}
\label{ads}
\end{eqnarray}
where the Dirac trace has been taken, and where these expressions are also to be understood in the sense of dimensional regularization.

Comparing these expressions for the coincident-point Green's functions (\ref{g}) and (\ref{ads}) for the gauge and gravity backgrounds, we see a strong similarity. To make this more explicit, recall that the one-loop effective Lagrangian can be deduced from the coincident-point Green's function as
\begin{eqnarray}
{\mathcal L}_{\rm gauge}&=&-\langle x | \ln (-D^2+m^2) |x\rangle =-\int^{m^2} G(x, x)
\label{gaugel}\\
{\mathcal L}_{\rm gravity}&=& \langle x | \ln (D \hskip -7pt /+m) |x\rangle =\int^{m} {\mathcal D}(x, x)
\label{ls}
\end{eqnarray}
This procedure is very well known so we do not repeat all the details here \cite{Candelas:1975du,Burgess:1984ti,gvd}, but as an illustration we recall the gauge case in $d=4$. An integral representation of the finite renormalized effective Lagrangian can be obtained by integrating the first line of (\ref{g}) with respect to $m^2$, as in (\ref{gaugel}). In $d=4$ this requires two subtractions, corresponding to the subtraction of the free field ($f=0$) Lagrangian and to charge renormalization \cite{gvd}: 
\begin{eqnarray}
{\mathcal L}_{\rm gauge}^{(d=4)}&=&\left(\frac{f}{4\pi}\right)^{2}\int_0^\infty \frac{ds}{s}\, e^{-m^2 s}
\left[\frac{1}{\sinh^{2}(f\, s)}-\frac{1}{(f\, s)^2}+\frac{1}{3}\right]
\label{4dg}
\end{eqnarray}
An alternative approach that is closer to that often used for gravitational effective actions \cite{Candelas:1975du,Dowker:1975xj,Allen:1983dg,Candelas:1983ae,Allen:1985wd,Burgess:1984ti,Camporesi:1992tm,odintsov}  is to expand the third line of (\ref{g}) about $d=4$,
\begin{eqnarray}
G(x, x)=\frac{m^2}{8\pi^2 (d-4)}+\frac{m^2}{16\pi^2}\left[\psi\left(\frac{m^2}{2f}\right)+\frac{f}{m^2}+\ln\left(\frac{f}{2\pi}\right)+\gamma-1\right]+O(d-4)
\label{expansion}
\end{eqnarray}
and then integrate with respect to $m^2$. In (\ref{expansion}) we see the appearance of the digamma function $\psi(y)\equiv\Gamma^\prime(y)/\Gamma(y)$, coming from the expansion of the gamma functions in (\ref{g}). After dropping the pole term and integrating the second term with respect to $m^2$, we  obtain the renormalized, finite effective Lagrangian. The resulting expression looks different from (\ref{4dg}), but in fact (\ref{4dg}) can be written as  \cite{gvd}:
\begin{equation}
{\mathcal L}_{\rm gauge}^{(d=4)}=\left(\frac{f}{2\pi}\right)^{2}\left[-\int_0^{m^2/2f} y\left(\psi(y)+\frac{1}{2y}-\ln y\right)dy +\zeta^\prime(-1)-\frac{1}{12}\ln \left(\frac{m^2}{2f}\right)\right]
\label{4dg2}
\end{equation}
Such expressions, obtained after different renormalization schemes, differ only up to some logarithm and polynomial terms which are remnants of the renormalization procedure and do not carry any special significance for our discussion. The physically relevant part is the term involving the digamma function $\psi(y)$. 
In dimensions higher than $4$, scalar QED is of course not renormalizable, but we define a natural finite effective Lagrangian by subtracting enough terms in the Taylor expansion of $1/\sinh^{d/2}(f s)$ in (\ref{g}) and (\ref{gaugel}) to make the proper time $s$ integral  convergent, a standard mathematical technique for defining a finite  determinant of a differential operator \cite{simon}, and which can also be understood naturally in terms of zeta function regularization \cite{elizalde}. These subtractions produce polynomial and log terms that correspond to renormalization scheme terms in $d=4$, and to the definition of the determinant for higher $d$, in the sense of \cite{simon}. In fact, the key part of the expressions  (\ref{4dg2}) and (\ref{expansion}) is the digamma function $\psi$, with all other log and polynomial terms being reconstructed straightforwardly from this. Repeating this procedure in $d=4n$ dimensions we obtain a simple expression for the effective Lagrangian as an integral of $\psi(y)$ times a special polynomial of $y$:
\begin{eqnarray}
{\mathcal L}_{\rm gauge}=-\frac{1}{\Gamma(d/2)} \left(\frac{f}{2\pi}\right)^{d/2} \int_0^{m^2/2f}dy \, y\, \prod_{k=1}^{d/4-1} (y^2-k^2) \,\psi(y)+\dots
\label{gauge-lag}
\end{eqnarray}
where "$+\dots$" denotes the aforementioned polynomial and log terms. 
Then the  effective action for a hypercube volume of side $L$ is
\begin{eqnarray}
{\mathcal S}_{\rm gauge}=-\frac{\left(\frac{f L^2}{2\pi}\right)^{d/2}}{\Gamma(d/2)}  \int_0^{m^2/2f}dy \, y\, \prod_{k=1}^{d/4-1} (y^2-k^2) \,\psi(y)+\dots
\label{gauge-lag-2}
\end{eqnarray}
Up to another irrelevant polynomial term, we then define the finite log determinant as the multiple gamma function \cite{Vardi,Adamchik:2003jc}:
\begin{eqnarray}
{\mathcal S}_{\rm gauge}=(-1)^{d/2} \left(\frac{f L^2}{2\pi}\right)^{d/2} \ln\Gamma_{d/2}\left(\frac{d}{4}+\frac{m^2}{2f}\right)
\label{gauge-multiple}
\end{eqnarray}
The prefactor  is the total flux. 

Repeating this procedure for a spinor field of mass $m$ in $d$ dimensional anti-de Sitter  (AdS) space [with $d$ even], represented by its Euclidean section $H^d$, we obtain the well-known effective Lagrangian  \cite{Allen:1985wd,Burgess:1984ti,Camporesi:1992tm,Avramidi:2000bm} from the coincident propagator trace (\ref{ads}):
\begin{eqnarray}
{\mathcal L}_{\rm gravity}=(-1)^{d/2-1}\frac{2}{\Gamma(d/2)} \left(\frac{\Lambda}{2\pi}\right)^{d/2} \int_0^{m/\sqrt{\Lambda}}dy \, y\, \prod_{k=1}^{d/2-1} (y^2-k^2) \,\psi(y)+\dots
\label{gravity-lag}
\end{eqnarray}
Multiplying by a volume element $\frac{\pi^{(d+1)/2}R^d}{\Gamma((d+1)/2)}$ of $H^d$,  with radius $R$, we obtain \cite{Burgess:1984ti}
\begin{eqnarray}
{\mathcal S}_{\rm gravity}=(-1)^{d/2-1} \frac{\left( 2\Lambda R^2\right)^{d/2}}{\Gamma(d)}\int_0^{m/\sqrt{\Lambda}}dy \, y\, \prod_{k=1}^{d/2-1} (y^2-k^2) \,\psi(y)+\dots
\label{gravity-lag-2}
\end{eqnarray}
Again,  this can be expressed in terms of the multiple gamma function  \cite{Vardi,Adamchik:2003jc,Das:2006wg} as 
\begin{eqnarray}
{\mathcal S}_{\rm gravity}=(-1)^{d/2}\left(2\Lambda R^2\right)^{d/2} \ln\Gamma_d\left(\frac{d}{2}+\frac{m}{\sqrt{\Lambda}}\right)
\label{gravity-multiple}
\end{eqnarray}
As before, the factor out front is the total flux. This computation is completely analogous to the mathematical definition of the determinant of the Laplacian on a  sphere, where the answer is also expressed in terms of multiple gamma functions \cite{Vardi,Dowker:2003ra,Das:2006wg}.

Our main observation is that the gauge effective action (\ref{gauge-multiple}) and the gravity effective action (\ref{gravity-multiple}) are expressed in terms of exactly the same function, a multiple gamma function, with the identifications
\begin{eqnarray}
{\rm gauge} \qquad &\leftrightarrow & \qquad {\rm gravity}\nonumber\\
d=4n \qquad &\leftrightarrow & \qquad d=2n \nonumber\\
{\rm massive\,\,scalar}  \qquad &\leftrightarrow & \qquad {\rm massive\,\,spinor} \nonumber\\
{\rm maximally\,\, symmetric\,\, gauge\,\, curvature} \qquad &\leftrightarrow & \qquad {\rm maximally \,\,symmetric\,\, gravitational\,\, curvature} \nonumber\\
\frac{m^2}{2f}  \qquad &\leftrightarrow & \qquad \sqrt{\frac{m^2}{\Lambda}}
\label{id}
\end{eqnarray}

Since the effective actions are the generators of one-loop scattering amplitudes with an arbitrary number of external photon or graviton (respectively) legs, this identification implies a nontrivial relationship between such amplitudes, with the gravitational curvature $\Lambda$ identified with the {\it square} of the gauge curvature $f$. This relation is valid in the low energy limit, since the effective actions have been computed for constant background fields. In the gauge theory case, the extraction of the explicit low-energy scattering amplitudes from the effective action is itself a nontrivial combinatorial problem \cite{Itzykson:1980rh,Dunne:2002qf,Martin:2003gb}, and in $d=4$ corresponds to diagrams with all external photons being of definite helicity [note these do not vanish, since the internal loop is massive]. We are not aware of any such explicit extraction of the low energy limit of graviton scattering amplitudes from the curved space effective action, although some steps have recently been taken in related directions \cite{Bastianelli:2008cu,Goncalves:2009sk,Fucci:2009je}.

To conclude, our basic observation in this short note is the equivalence of the gauge theory effective action (\ref{gauge-multiple}) and the gravity effective action (\ref{gravity-multiple}), subject to the identifications (\ref{id}). But the essential suggestion is that it may be fruitful to explore the gauge-gravity scattering amplitude relation in the language of the generator of scattering amplitudes, namely the effective action. There are several natural directions in which our observation might be extended. First, one could relax the condition of maximal symmetry, in which case the gauge theory effective action can be expressed in terms of the generalized Barnes multiple zeta and gamma functions \cite{ruijsenaars}, while the gravitational effective action can be studied on more general symmetric spaces \cite{Avramidi:2007rh}. Second, the constant background field can be generalized to inhomogeneous background fields using the Fock-Schwinger gauge expansion for the gauge theory, and the Riemann normal coordinate expansion for the gravitational background. Another interesting question is whether anything simple happens beyond the one loop level, for example in a maximally supersymmetric setting. On the gauge side, a miraculous recurrence relation between the one-loop and two-loop effective action is known to exist for these maximally symmetric background fields, which are self-dual in 4 dimensions \cite{Dunne:2002qf}, and hence preserve supersymmetry. Similar dramatic simplifications also occur at two-loop for the effective action in supersymmetric theories such as SQED and $N=4$ SYM \cite{Kuzenko:2003qg}.
It would be interesting to know whether  anything similar might happen on the gravity side.

\bigskip

\bigskip

We thank the US DOE for support through grant DE-FG02-92ER40716. 

\section{Appendix: Multiple Gamma Functions}
\label{mg-sec}

The multiple gamma functions $\Gamma_n(z)$ were introduced long ago by Barnes \cite{ruijsenaars}, and have since been understood as the natural functions to describe determinants of operators on spheres \cite{Vardi,Dowker:2003ra,voros,quine}. They can be defined uniquely \cite{vigneras} by the conditions:
\bea
\Gamma_{n+1}(z+1)&=&\frac{\Gamma_{n+1}(z)}{\Gamma_n(z)}
\label{mg1} \\
\Gamma_1(z)&=&\Gamma(z)
\label{mg2}\\
\Gamma_n(1)&=&1
\label{mg3}\\
(-1)^{n+1}\frac{d^{n+1}}{dz^{n+1}} \ln \Gamma_n(z)&\geq& 0
\label{mg4}
\eea
Various integral representations and asymptotic expansions can be found in \cite{ruijsenaars,Burgess:1984ti,Adamchik:2003jc}. In some papers these functions are written in terms of $G_n(z)$ where 
$\Gamma_n(z)=\left[G_n(z)\right]^{(-1)^{n+1}}$. The most useful representation for our purposes can be derived from a result listed in \cite{Burgess:1984ti}:
\bea
\ln \Gamma_n(1+z)=\frac{(-1)^{n+1}}{(n-1)!}\int_0^z dx\, \left[\prod_{j=0}^{n-2}\left(x-j\right)\right]\,\psi(1+x)+{\mathcal Q}_n(z)\quad ,
\label{mg-integral}
\eea
where the ${\mathcal Q}_n(z)$ are known polynomials of degree $n$. Noting that this is also written as an integral of a polynomial times the digamma function $\psi$, it is straightforward to derive the expressions (\ref{gauge-multiple}) and (\ref{gravity-multiple}) for the log determinants in terms  of multiple gamma functions from (\ref{mg-integral}). The additional polynomial terms arising in these manipulations are collected into the definition of the regularized determinant, just as is done for the sphere \cite{quine}.

\end{document}